\documentclass[12pt,oneside]{amsart}

\usepackage{amsfonts}
\usepackage{amssymb}
\usepackage{amsmath}
\usepackage{amsthm}

\usepackage{graphicx}

\begin{document}

\title[]{Spreading dynamics on spatially constrained complex brain networks}

\author{Reuben O'Dea$^{1}$}
\address{$^1$School of Science and Technology, Nottingham Trent University, Nottingham, NG11 8NS, UK} 
\author{Jonathan J. Crofts$^{1}$}

\author{Marcus Kaiser$^{2,3,4}$}
\address{$^2$School of Computing Science, Newcastle University, Newcastle upon Tyne, NE1 7RU, UK}
\address{$^3$Institute of Neuroscience, Newcastle University, Newcastle upon Tyne, NE2 4HH, UK}
\address{$^4$Department of Brain and Cognitive Sciences, Seoul National University, South Korea}

\date{\today}


\begin{abstract}
The study of dynamical systems defined on complex networks provides a natural 
framework with which to investigate myriad features of neural dynamics, and has 
been widely undertaken. Typically, however, networks employed in theoretical 
studies bear little relation to the spatial embedding or connectivity of the 
neural networks that they attempt to replicate. Here, we employ detailed 
neuroimaging data to define a network whose spatial embedding represents 
accurately the folded structure of the cortical surface of a rat and investigate 
the propagation of activity over this network under simple spreading and 
connectivity rules. {By comparison with standard network models with the same 
coarse statistics, we show that the cortical geometry influences profoundly the speed
propagation of activation through the network. Our conclusions are of high 
relevance to the theoretical modelling of epileptic seizure events, and indicate 
that such studies which omit physiological network structure risk simplifying the 
dynamics in a potentially significant way.}
%
\end{abstract}

\maketitle

\section{Introduction}\label{sec:intro}
The newly emerging discipline of Network Science provides a general framework 
for representing, modelling and predicting the behaviour of complex systems 
belonging to areas as diverse as social science, biology and information 
technology \cite{newman10}. Motivated by the observation that most real-world 
networks fail to conform to the homogeneous Poissonian degree distribution 
admitted by Erd\H{o}s-R\'{e}nyi random graphs \cite{erdos59}, improved network 
models were constructed (most notably the {\it small-world model} of Watts and 
Strogatz \cite{strogatz98} and the {\it preferential attachment model} of 
Barabasi and Albert \cite{barabasi99}) that were capable of recovering many 
of the interesting features displayed by real-world network data. Initial 
investigations into complex networks focussed primarily on the characterisation 
of networks in terms of a small number of topological parameters; however, more 
recently, interest has shifted towards understanding the influence of network 
structure on the dynamic processes occurring upon them -- see, for example, 
\cite{arenas08,barrat08} and references therein. 

Such investigations are particularly relevant to biological systems, in which 
a well-defined network structure is frequently a key feature, the evolution
and topology of which are presumed to affect relevant biological processes. A paradigm 
for such studies is that of neural systems, which have been widely studied in 
this context with the aim of providing a more complete understanding of epilepsy 
and other neural conditions \cite{sporns10}. Epilepsy is characterised by recurrent 
and unpredictable instances of ``excessive, or synchronous neuronal activity'' 
(seizures) \cite{fisher2005}; the synchronisation of neuronal activity in networks 
has therefore received significant attention, and network topology is considered 
to be a dominant factor affecting spreading dynamics 
\cite{deville12,kaiser10,kaiser07,Kotter2000,roxin2004self,Kramer2010}, independent of the specific  
model by which activation is transmitted across the network.  A wide range of 
transmission models has been employed in the literature. Representative examples 
include simple cellular automata-type spreading rules 
\cite{kaiser07,newman2005power} and pulse-coupled-oscillators \cite{deville12} 
(and references therein), of which integrate-and-fire models
\cite{roxin2004self,kaiser12pre,netoff2004epilepsy} are a special case. 
We note that in, {\it e.g.}~\cite{deville12,kaiser10,kaiser07} (and in 
the present work), network nodes are to be interpreted as `neural units' 
comprising many neurons, and representing a single cortical column, 
say. Detailed consideration of synaptic signalling models is therefore not 
appropriate.

A further essential aspect of neural systems is  that they are spatially 
embedded -- {\it i.e.},~their nodes and edges are constrained to lie on a  fixed 
geometric structure, and it is expected that these additional factors will further  
influence network dynamics. Indeed, though spatial networks have received considerable attention of  late \cite{barthelemy11}, the influence of spatial embedding upon network dynamics is not well understood. 

{A small number of recent studies have begun to employ neuroimaging data ({\it e.g.}~via Diffusion Tensor Imaging) to infer large-scale network connectivity \cite{Sotero2007,Taylor2012}; however, the majority of studies in macroscale epilepsy and seizure modelling typically employs uniform lattices in 1D or 2D 
with network connectivity restricted to nearest-neighbours, or with additional long-range connections obeying arbitrary rules \cite{goodfellow12,netoff04,rothkegal09}. Such an approach fails to exploit a wealth of neuroimaging data \cite{kaiser11}, which reveals the intricate connectivity structure and surface geometry of the brain.} Therefore, current approaches still observe simplified 2D surfaces \cite{Winfree1994,Kim2009,Traub} inspired by the original studies of excitable media \cite{Wiener1946} --- an organization that is different from the rounded shape of model organisms such as  rodents or convoluted brain surfaces such as for humans. Note that this limitation not only relates to epilepsy modelling but also to studies of spreading depression \cite{Tuckwell1978,Dahlem2008}.

{In this work, we seek to address the deficiency in neural network studies discussed above.} We employ detailed spatial information obtained from freely-available neuroimaging data of a rat brain to define a physiologically-relevant neural network and, via comparison with commonly-employed network architectures, determine the influence of the connectivity and complex surface geometry of the brain on seizure dynamics. We restrict attention to lattice-like network structures ({\it i.e.}, networks that display significant clustering and long average path-lengths), and a basic spreading model in favour of the more complex signalling models outlined above, as the aim here is to highlight the contribution of spatial network properties to signal transmission. We also limit our current study to spreading on the brain surface; the role of long-distance white matter fibre tracts between brain regions is not part of this work. 

{Our investigations highlight clearly that employing cortical surface geometry to inform network structure influences dramatically the propagation of activation through the network. By doing so, we indicate that activation dynamics of relevance to epileptic seizure initiation and progression (in particular, those highlighting the importance of the site of initiation within the network) display distinct differences in idealised networks which are commonly employed in the literature. Most strikingly, we show that our cortical network delays significantly the total activation of the network when compared to idealised networks with the same coarse statistics: in the parameter regimes that we study, the time to activation is increased by a delay factor lying in the range 1.45--1.88. In this way, we highlight clearly the importance of realistic network structure on activation dynamics and indicate that such considerations should be included in theoretical models which aim to provide a more complete description of ({\it e.g.}) epileptic seizure activity.}
 
The remainder of this paper is organised as follows. In \S\ref{sec:spatial}, we 
use neuroimaging data to construct a lattice-like network architecture embedded on 
the cortical surface of a rat brain, together with a simple cellular automaton-like 
rule governing network activity. In \S\ref{sec:res}, we present a comparative 
analysis of spreading dynamics over the cortical network, a uniform square lattice 
and an ensemble of 2D geometric random graphs.  A summary of our main results, together with a discussion of future avenues of investigation, is provided in \S\ref{sec:conc}.

\section{Spatial complex networks}\label{sec:spatial} 

In this paper, we investigate the influence of network structure on the dynamics 
of the processes occurring upon them, with specific application to neural 
signalling. We achieve this by comparing the spreading dynamics of simulated 
neural activation within a rat cortex with those obtained on commonly-employed 
spatial networks. All of the networks examined herein are unweighted, undirected, 
and without loops. 


A plethora of tools and techniques for characterising complex networks exists \cite{costa07}; however, in the context of spatial networks not all of these measures remain relevant. An important feature of neural networks is the propensity for nearby nodes to connect, and thus, such graphs tend to exhibit high levels of local clustering. In addition, vertex reachability (defined as the ability to travel between nodes $i$ and $j$ following connections within the network) impacts significantly on spreading dynamics. {In view of this, systematic  comparison of the network models considered here will be effected via the {clustering coefficient}, which is given mathematically as   
\begin{equation}
 C = \frac{1}{N}\sum_{i=1}^{N}{C_i},
\end{equation}
where $C_i$ denotes the probability that any two neighbours of node $i$ are connected and $N$ the order of the graph. Additionally, we employ the {characteristic path-length}, $L$, which is defined as the number of edges in the shortest path between two vertices, averaged over all pairs of vertices. We remark that if the network is not fully connected then the characteristic path-length diverges, as a disconnected pair corresponds to an infinite path.}


\begin{figure}[htbp!]
\includegraphics[width=\textwidth]{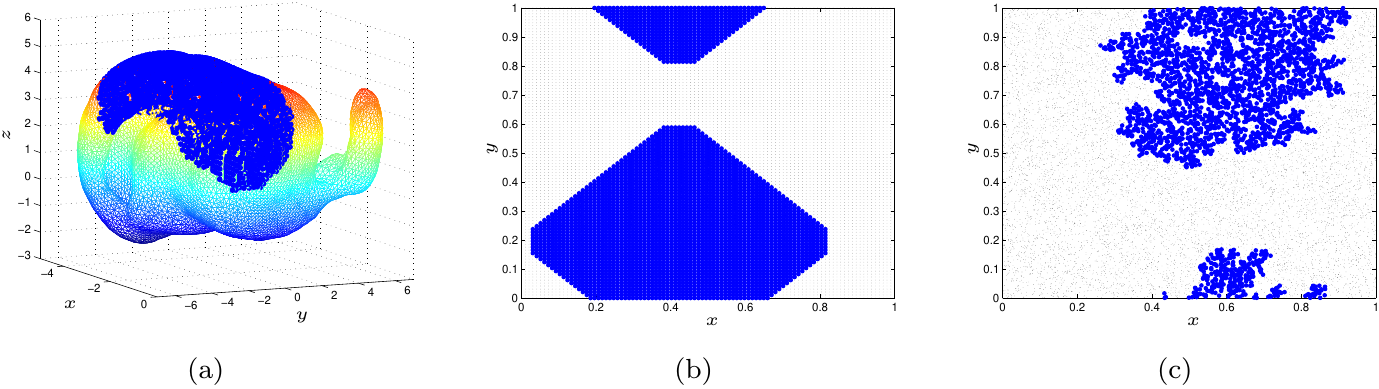}
\caption{A typical simulation of spreading dynamics in spatially constrained networks. Blue dots denote activated nodes (colour online). (a) A spatial network embedded on the left-hemisphere of the rat cortex; (b) a 2D square lattice graph; (c) a 2D periodic random geometric graph.}
\label{fig:sample_activation}
\end{figure}

\subsection{Rat cortical network}\label{subsec:rat}

Spatial coordinates defining the cortical surface of the rat  were obtained from 
the Caret software package \cite{vanessen12} and processed using the Caret Matlab toolbox. Figure \ref{fig:sample_activation}(a)  shows the cortical surface of the left hemisphere of the rat brain, with typical neural activity spreading obtained via numerical simulation (see \S\ref{sec:res}) superimposed.

{Restricting to the left hemisphere, we construct a cortical rat network, with nodes positioned on the $N_{\mathrm{rat}}=9623$ available data points (see Figure 
\ref{fig:sample_activation}(a)), by connecting nearest neighbours according to 
the following process: (i) a minimally-connected nearest-neighbour network is defined via the triangulation provided by the Caret software package; (ii) vertex pairs are connected if they lie within a Euclidean distance $r$ of each other; (iii) connections are removed if the shortest path, calculated using the nearest-neighbour network defined in (i), between connected nodes exceeds a predefined number of steps, which is defined experimentally.  While Euclidean distance is a reasonable estimate of physiological connection length in general \cite{kaiser11}, the third step is necessary to remove spurious edges which arise for large $r$, that are near in the ambient space, yet distant as measured on the cortical surface: shortest path length, defined via a simple mesh triangulation provides a simple method with which to measure this disparity.}

%

\subsection{Standard network models}\label{subsec:standard}


To discern the importance of network structure on activation spreading dynamics, 
we compare the spreading dynamics observed on the cortical network defined in \S\ref{subsec:rat} with that observed on a uniform square lattice graph and on {an ensemble of two-dimensional geometric random graphs \cite{penrose05},} network structures typical of those employed in the literature \cite{goodfellow12,netoff04,rothkegal09}.

To construct a random geometric graph we place uniformly and independently $N$ 
nodes at random on the unit square, and form connections between pairs of nodes 
according to Euclidean distance. Similarly, a square lattice graph is obtained by forming distance-dependent connections within a $N$-point uniform square lattice on the unit square. Figures \ref{fig:sample_activation}(b,c) illustrate these networks, and a typical simulation of neural activity.

For comparability with the cortical network outlined in \S\ref{subsec:rat}, we choose $N=9604$ (the closest square number to $N_{\mathrm{rat}}$) and control the number of edges in the graph via $\tilde{r}$, the Euclidean distance scaled on the surface area of the rat cortex. The surface area is calculated via the Caret Matlab toolbox to be $S\approx221$ (arbitrary units), which provides a scaling: $\tilde{r} = r/\sqrt{S}$. Additionally, we impose periodic boundary conditions.

\begin{figure}[htbp!]
\includegraphics[width=0.8\textwidth]{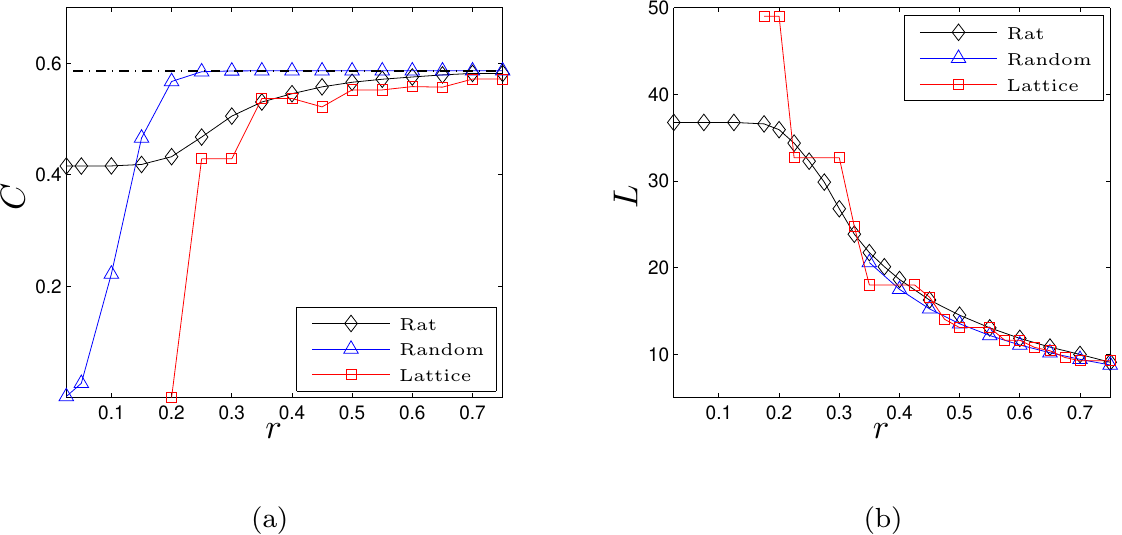}
\caption{Network measures. The clustering coefficient $C$ and average path length 
$L$ of each network plotted as a function of the connectivity distance, $r$. The dot-dashed line in (a) indicates the asymptote provided by Equation (\ref{gammafunctionalwaysnice}). Data for the random graph is calculated from 10,000 realisations; in the parameter range $r\in[0.35,0.5]$ the maximum standard deviation of $C$ and $L$ is $\sigma = 0.0016$ and $0.0301$, respectively.}
\label{fig:networkmeasures}
\end{figure}

Figure \ref{fig:networkmeasures} shows how the network measures $C$ and $L$, with 
which we quantify differences between the three networks, vary with the connectivity distance, $r$. Ensemble measures for the random graph are calculated from 10,000 realisations.  These results indicate that the networks exhibit distinctly different features when restricted to short-range connections (small $r$), while highly-connected networks are comparable.
{It is noteworthy that the value of the mean clustering coefficient for the random graphs tends to the theoretical value}
\begin{equation}
	1-\frac{1}{\Gamma{\left(\frac{3}{2}\right)}\sqrt{\pi}}
	\left(\frac{3}{4}\right)^{\frac{3}{2}} \approx   0.58650\label{gammafunctionalwaysnice}
\end{equation}
{obtained in \cite{dall02}, where $\Gamma{\left(x\right)}$ denotes the 
gamma function.}

We remark that for $r<0.2$ we find $\tilde{r}<1/\sqrt{N}$, and therefore no connections exist in the square lattice graph (see Figure \ref{fig:networkmeasures}(a)). At $r=0.2$ the connectivity of the lattice graph corresponds to a four-point nearest-neighbour stencil (so that the connectivity matrix is analogous to a discrete Laplacian operator), and since nearest-neighbours of a vertex are not neighbours of each other, it follows that $C=0$. {In addition, note that due to interplay between the regular discrete structure of the lattice and our rule based upon Euclidean distance with which to add new edges, variation of the clustering 
coefficient for the lattice graph is non-monotone with respect to $r$.} The mean degree of the other networks is $d\approx 5$ (data not shown). 
Lastly, we note that for $r<0.35$, the random graphs are not fully-connected, hence the absence of data in Figure \ref{fig:networkmeasures}(b).

\section{Spreading dynamics}\label{sec:res}

Propagation of activation within each neural network defined in \S\S\ref{subsec:rat} 
and \ref{subsec:standard} was governed by a basic spreading model 
\cite{kaiser07,newman2005power}, summarised as follows. 

Nodes $i$ are restricted to exist in one of two states: active ($x_i=1$), or inactive 
($x_i=0$). Starting from an initial activation state, simulation operated in discrete timesteps; from one timestep to the next, an inactive node became activated (or an active node remained in the active state) if it was connected to at least $m$ active nodes.  {Initial conditions comprised a small region of activation (1$\%$ of the total nodes in the network) surrounding a node selected at random; the propagation of this activation through the network under our simple assumptions on spreading dynamics provides a convenient and compelling method with which to highlight the differences imparted by the networks under consideration. We 
choose the mean fraction of activated nodes as our key metric with which to investigate the different networks. Ensemble measures of network dynamics on the 
random graphs and rat network were constructed from 10,000 simulations; behaviour in the uniform lattice is identical for all initial activation positions.}



{The value of $m$ places a lower bound on the connectivity of the network for which network activation can occur, and influences the speed of spreading of activation in the network (and, together with the value of $r$, the shape of the advancing activation front). Since we consider highly simplified dynamics in this study, omitting, for example, random inactivation or complex intra-node dynamics (the better to emphasise the importance of network structure on activation dynamics), the balance between $m$ and $r$ determines completely the speed of activation of the network (indeed, for appropriate $m$ and $r$,  whole network excitation is inevitable) and, furthermore, affects all networks in the same manner. In all of the simulations that follow we chose $m=2$ without loss of generality.} { For $r<0.35$, the network structures are not comparable (as highlighted by Figure \ref{fig:networkmeasures}(b)); for $r>0.5$, the networks are highly-connected and activation spreads rapidly over the cortical surface. Our interest here is in the short-range connections between cortical columns, rather than long-range white matter fibre tracts, say. We therefore restrict attention to connection distances $r\in[0.35, 0.5]$, leading to mean degrees in the cortical network $\langle d_{\mathrm{rat}}\rangle\in [17.59, 37.36]$. 
{Whilst actual estimates for mean degree in the rat cortex are not available, such connectivity is typical of that employed in the literature (see, 
{\it e.g.}~\cite{deville12,kaiser10,kaiser07} and references therein)}, particularly when one accounts for the large spread of observed degrees ({\it e.g.,} for $r=0.35$ the degree ranges from $8$ to $48$ whilst for $r=0.5$ it lies between $20$ and $87$), and so serves to illustrate our methodology. For brevity, in the figures that follow, we illustrate the differences in spreading dynamics obtained in each network for the choices $r=0.35$ and $r=0.5$.}


{Figure \ref{fig:sample_activation} shows a typical pattern of activation at an illustrative timepoint in each of the three networks, which serves to highlight clearly how differences in the underlying network connectivity are made manifest in the spreading of activation through the network. We remark that patterns of excitation are `well-defined' and uniform in the rat and 
lattice graphs, with all nodes within a region activated, while in the random graphs, lack of connectivity can lead to bottlenecks (here, we choose $r=0.25$ to highlight these structural differences).}

\begin{figure}[htbp!]
\includegraphics[width=0.8\textwidth]{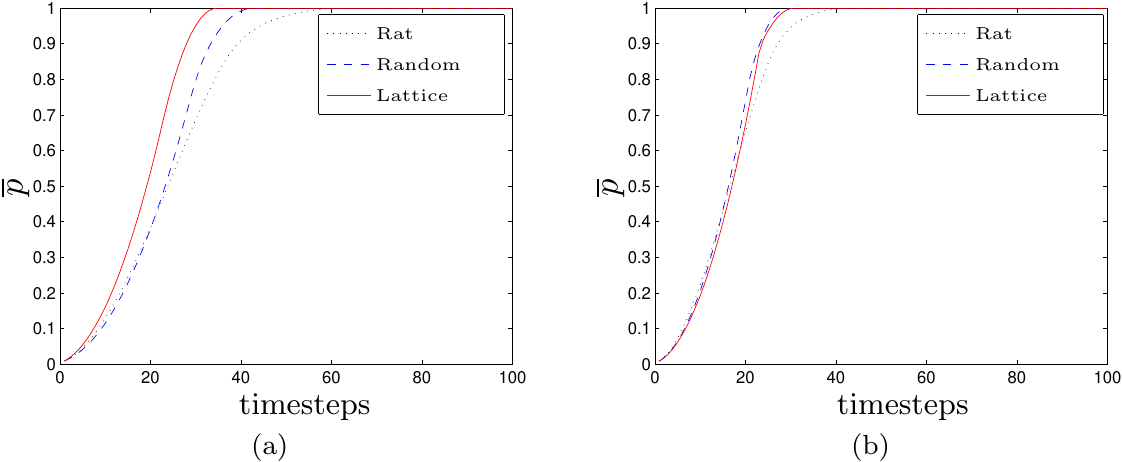}
\caption{The evolution of the mean fraction of activated nodes, $\overline{p}$, in 
the rat network, random graphs and lattice for different connectivity distances: 
(a) $r=0.35$, (b) $r=0.5$.}
\label{fig:meanfraction}
\end{figure}

Figure \ref{fig:meanfraction} shows the spread of activation through each network, as measured by the fraction of activated nodes (averaged over all instances).  Figure \ref{fig:meanfraction}(a) indicates that for lower network connectivity ($r=0.35$; the mean degree in each network was comparable: 
$\langle d_{\mathrm{rat}}\rangle=17.59$, $\langle d_{\mathrm{rand}}\rangle=16.09$ and $d_{\mathrm{latt}}=20$)
the activation speed in each of the three networks differs significantly: the lattice graph requires dramatically fewer timesteps to achieve entire network activation; the cortical network is the slowest of the three. In fact, we observe a 1.88-fold and 1.45-fold increase in activation time in the rat network when compared to the lattice and random graphs, respectively. For more highly-connected networks ($r=0.5$), the cortical network remains the slowest to activate; however, the lattice and random graphs now display similar activation rates, with the uniform lattice being marginally faster. The delay imparted by the cortical network is now 1.65-fold (lattice) and 1.46-fold (random graphs).

\begin{figure}[t]
\includegraphics[width=0.9\textwidth]{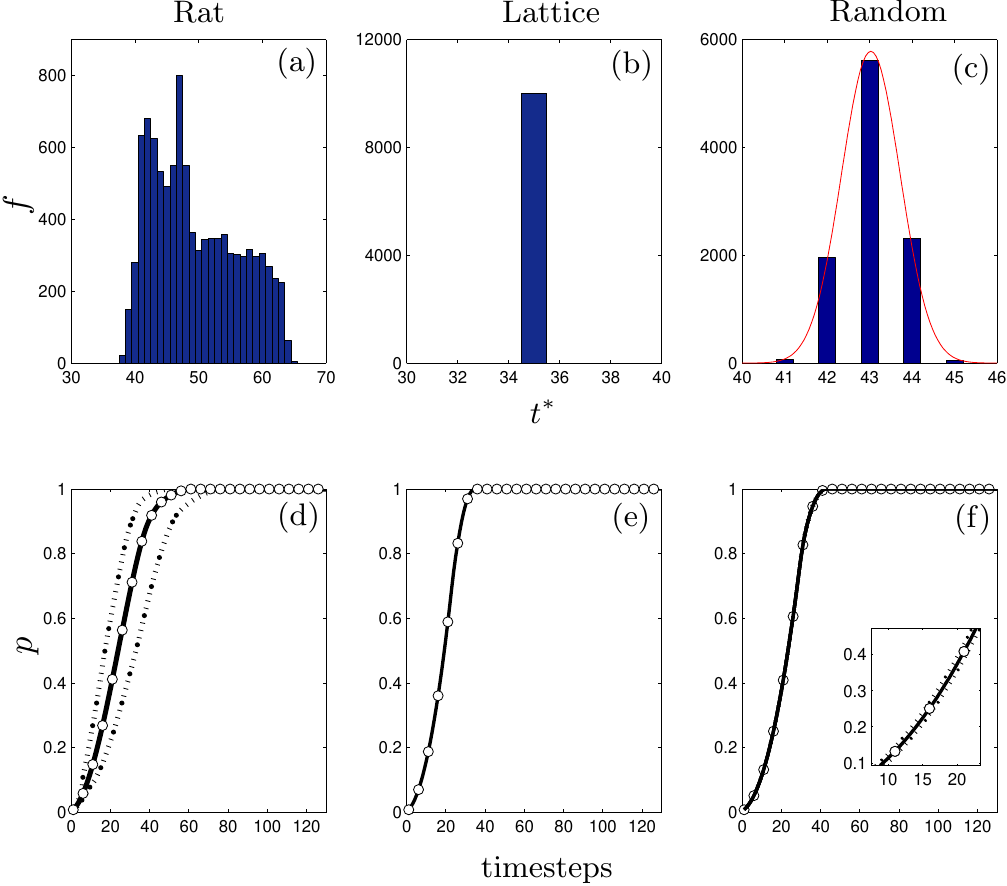}
\caption{Simulation results indicating the variability in activation dynamics in each network. Panels (a), (d) contain results in the cortical network; (b), (e), the lattice; and (c), (f) the random graphs. Upper panels (a)--(c): Histograms showing the time to full activation $t^*$. In (c), a comparison with a Gaussian distribution ($\mu=43.03$, $\sigma=0.6901$) is also shown. Lower panels (d)--(f): The evolution of the mean fraction of activated nodes in each network (white circles), together with a confidence interval of width $2\sigma$. The connectivity distance is chosen 
as $r=0.35$.}
 \label{fig:initialconds}
\end{figure}

\begin{figure}[htbp!]
\includegraphics[width=0.9\textwidth]{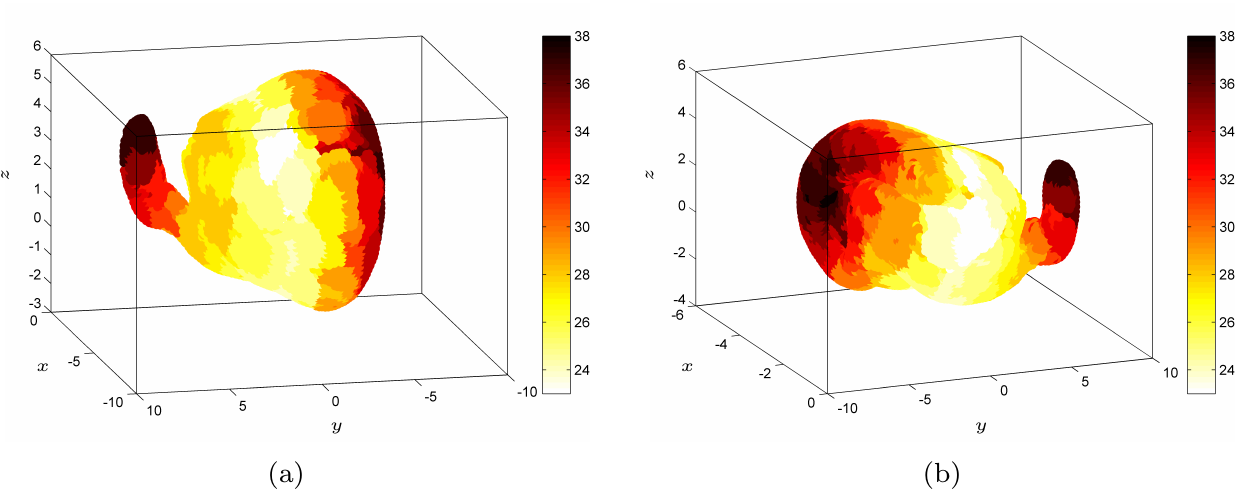}
\caption{Heat maps indicating the dependence of activation dynamics on the initial activation site. The colours of each region indicate the time to full network activation, $t^*$, associated with initial activation in that region (colour online). Panels (a) and (b) show rotated views; in each case $r=0.5$.}
\label{fig:heatmap}
\end{figure}

Figures \ref{fig:initialconds}  and \ref{fig:heatmap} highlight the influence of initial activation position  on the spreading dynamics, which is quantified by the time to full-network activation, $t^*$. 
For the values of connectivity distance, $r$, analysed here, full network spreading was observed for all networks independent of topology. Therefore, all simulations contributed to the calculation of $t^*$.

In Figures \ref{fig:initialconds}(a)--(c) histograms are presented, depicting 
the spread of $t^*$ observed in the simulations for each network; Figures 
\ref{fig:initialconds}(d)--(f) show the corresponding activation spread for each simulation, together with the ensemble mean.  {The delta-function obtained in Figure \ref{fig:initialconds}(b) and the corresponding results shown in Figure 
\ref{fig:initialconds}(e), reflect the fact that the network dynamics are identical for all realisations}, due to the uniform structure throughout the lattice, and the periodic boundary conditions. In contrast, Figures \ref{fig:initialconds}(a) and (d) indicate a very wide spread of activation times, and with no clear distribution, in the cortical network. {In the random graphs, $t^*$ displays small variation on each realisation; a wider spread is exhibited for a less well-connected network (data not shown). As highlighted by Figure \ref{fig:initialconds}(c), this distribution is well-characterised by a Gaussian with mean $\mu=43.03$ and variance $\sigma=0.6901$ ($p<0.01$). The observed distributions of spreading dynamics for the cortical and the random networks} (Figures \ref{fig:initialconds}(a) and \ref{fig:initialconds}(c) respectively) were found to differ significantly ($p<10^{-5}$ over the investigated parameter range) according to the Kolmogorov-Smirnov statistic.

In Figure \ref{fig:heatmap} heat maps are presented to highlight those initial activation sites in the cortical network which provide significantly higher full-network activation speeds.  These results indicate that the differences in activation time shown in Figure \ref{fig:initialconds} are due to the geometric structure of the rat brain and the presence of folded and smooth regions in the rat cortical surface.

The results shown in Figure \ref{fig:heatmap} are for the highly-connected network ($r=0.5$); however, qualitatively similar results are obtained for the range of parameter values analysed here: although values of $t^*$ vary, in all cases, initial activation of the folded regions leads to (approximately) a 1.7-fold reduction in total of activation time (data not shown).

{We have presented simulation results which highlight the differing rates of network activation, from an initial state comprising a small region of activated nodes. Our model can therefore be thought of as broadly applicable to the initial stages of epileptic partial seizures, whereby spreading can initiate from a certain region, leading to increased activity patterns in larger parts of the brain 
(mechanisms that lead to the rise of initial activity, such as high frequency oscillations through gap junctions \cite{roopun2010}, or to seizure termination are not considered). Additionally, the results that we have presented, indicating the dependence on activation progression on initiation site are pertinent to observations in epilepsy, as not all brain regions have the same probability of 
being starting points for seizures. We remark, however, that the concept of a specific focal point of seizure origin from which seizure activity spreads (due to local abnormal connectivity) has been criticised (see 
\emph{e.g.}~\cite{
Spencer2002}), and alternative hypotheses presented \cite{Stead,Goodfellow}. We do not discuss this further as our focus here is on the global network structure imparted by the rat cortex embedding and its effect on activation propagation.}

{We remark that although our model has relevance to epileptic seizure spreading processes as described above, its form is intentionally simplistic: we have omitted a plethora of physiologically important network and signalling features (such as long-range connectivity, or complex node activation dynamics) as our study addresses a fundamental aspect of the theoretical modelling of neural networks. Our central result is that cortical surface geometry affects profoundly signal propagation through the network, when compared to idealised networks with the same coarse statistics: the results contained in this section highlight clearly that significant differences in spreading dynamics are obtained in the network based on the cortical structure of the rat brain, compared to more standard network models.} Since networks analogous to these standard models (defined in 
\S\ref{subsec:standard}) are employed frequently in the theoretical literature, in order to study the features of epileptic seizure initiation and dynamics, the simulation results highlighted by Figures \ref{fig:meanfraction}, \ref{fig:initialconds} and \ref{fig:heatmap} indicate clearly that certain dynamics of relevance to seizure initiation and progression (in particular, those highlighting the importance of the site of initiation within the network) are not well-captured by such approaches.  



\section{Discussion}\label{sec:conc}
In this paper, we have investigated numerically the influence of the structure of a 
spatially-embedded complex network on the dynamics of the processes which occur upon it, {with application to the development of improved theoretical models of the progression of epileptic seizures and spreading depression over the cortical surface.}

The key feature of this work is that we employ neuroimaging data from the left hemisphere of a rat brain to define a neural network whose spatial embedding represents accurately the structure of the cortical surface. Typically, theoretical studies of neural network dynamics employ uniform lattices or random graphs in 1D or 2D: their focus being on in-depth analysis of various theoretical connectivity rules ({\it e.g.}~small world or scale-free networks) or signalling processes on the network dynamics. Here, we define connectivity within our cortical network by a simple criterion based on Euclidean distance (modified to account for the folded cortical structure), and employ a basic spreading rule to govern the propagation of activation. In this way, we highlight clearly the influence of physiologically-relevant network structure on seizure dynamics, in isolation. 

We compared numerical simulation of the propagation of neural activity within three different network architectures: the cortical neural network, a uniform square lattice graph and an ensemble of two-dimensional geometric random graphs, and studied these over a range of network connectivities. { Note, that we 
limit our current study on comparisons of spreading on the brain surface and to short-range connections between nearby cortical columns; the role of long-distance fibre tracts between brain regions is not part of this work.}  In the parameter range chosen for the dynamic simulations which we investigate in detail, these networks were shown to be comparable in terms of their clustering coefficient and characteristic path length; in the case of networks with low connectivity 
(\emph{\it i.e.}~for values of $r$ approaching the uniform lattice spacing) the networks display distinctly different characteristics. Despite their comparability, however, our simulations indicate dramatic differences in the propagation of activation through the various networks: in particular, we observe that the time taken to full activation in the cortical network is increased by a factor lying in the range 1.45--1.88. Especially striking is the importance of the site of activity initiation: a dramatic spread in the number of time-steps 
required to achieve full network activation is observed in the rat network, even in the case of a highly-connected network ($r=0.5$, $\langle d_{\mathrm{rat}}\rangle=37.36$). Small variation is observed in the random graph; the dynamics on the lattice graph is independent of initiation site. For less well-connected random graphs ($r=0.35$, $\langle d_{\mathrm{rand}}\rangle=16.09$), 
the variation is increased; however, the differences between all three networks remain significant (data not included).  { Moreover, our simulations highlight that, across the parameter range investigated here, different initial activation sites lead to approximately a 1.7-fold reduction in full-network activation time, due to variation in connectivity across these regions.}


{We employ a highly simplified model with which to investigate the spreading of activation over our network. Various features of relevance to physiological neural networks, such as complex node dynamics, long-range connectivity, or stochasticity, are omitted from our formulation; indeed, our model draws no distinction between normal activation spreading, and that seen in epilepsy. However, such an approach allows us to provide a more powerful exposition of the importance of the network structure imparted by the cortical geometry, in isolation. We have shown that certain features of the activation dynamics display distinct differences, when compared to idealised models of the type commonly employed in the literature. Most strikingly, we highlight a variability in activation time for the cortical network, depending on initial activation site. This is pertinent to observations in epilepsy, as not all brain regions have the same probability of being starting points for seizures. Variability in cortical networks is well-studied in terms of structural and functional connectivity of brain regions, for example concerning the degree of nodes \cite{Eguiluz:2005uq,Achard:2006kx,Sporns:2007vn} and its consequence on network robustness and performance \cite{Kaiser:2007fk,Zamora-Lopez:2010ys}. However, our result is remarkable in that variability is observed for curved brain surfaces even in the absence of white-matter fibre-tract connectivity. Our results lead us to conclude that studies which do not take into account the spatial 
embedding of the cortex risk simplifying neural activation dynamics in a potentially significant way and are, therefore, unlikely to be able to represent accurately activation dynamics of relevance to epileptic seizures. We note, however, that in a physiological setting, similar disparity in network activation may be induced by 
including a range of other factors since the influences on the network dynamics within physiological neural networks are myriad; indeed, there is general 
agreement that no single factor can explain the varied phenomena associated with epileptic seizure dynamics.}




We remark that epileptic seizure events are extremely rare 
in wild-type rodents, and characteristics of epileptic activity in animal models can differ from those observed in humans \cite{roopun2010}. {However, our initial study employing a network based upon the cortex of a healthy rat has highlighted the potential importance of geometric structure in activation dynamics; similar investigations in a network whose spatial embedding represents the convoluted human 
cortical surface therefore forms important ongoing work.} In addition, future work will investigate how our predictions are altered under (i) a signal transmission model which represents more accurately the behaviour a neural unit and (ii) the 
introduction of long-range connections representing white-matter structural connectivity.


\section*{Acknowledgements}

M. Kaiser was supported by the WCU program through the KOSEF funded by the MEST (R31-10089), 
EPSRC (EP/G03950X/1), and the CARMEN e-science project (\texttt{http://www.carmen.org.uk}) 
funded by EPSRC (EP/E002331/1).


\end{document}